\documentclass[superscriptaddress,groupedaddress,nofootnoteinbib,12pt]{article}  

\usepackage{graphicx}  
\usepackage{dcolumn}   
\usepackage{bm}        
\usepackage{jcappub}
\usepackage{enumitem}

\def\be{\begin{equation}}
\def\ee{\end{equation}}
\def\ba{\begin{eqnarray}}
\def\ea{\end{eqnarray}}
\def\beq{\begin{eqnarray}}
\def\eeq{\end{eqnarray}}

\def\mpl{M_{\rm Pl}}
\def\e{{\epsilon}}

\def\d{\mathrm{d}}
\def\p{{\cal P}}

\def\L*{{\cal L}_*}
\def\L{\mathcal{L}}
\def\({\left(}
\def\){\right)}
\def\ie{{\it i.e. }}

\def\nn{\nonumber}
\def\p{\partial}

\def\stu{St\"uckelberg }
\def\p{\partial}

\def\<{\langle}
\def\>{\rangle}

\def\cs2{c_{s}^{2}}

 \def\ep{\varepsilon}

 \def\om{\omega}

 \def\p{\partial}

 \def\wed{\wedge}

 \def\be   {\begin{equation}}   \def\ee   {\end{equation}}

 \def\ba  {\begin{eqnarray}}   \def\ea  {\end{eqnarray}}

\begin{document}

\title{New Kinetic Terms for Massive Gravity \& Multi-gravity: A No-Go in Vielbein Form}
\author{Claudia de Rham, Andrew Matas and Andrew J. Tolley}
\affiliation{CERCA, Department of Physics, Case Western Reserve University, 10900 Euclid Ave, Cleveland, OH 44106, USA}
\date{\today}

\abstract{We reconsider the possibility of a class of new kinetic terms in the first order (vielbein) formulation of massive gravity and multi--gravity. We find that new degrees of freedom emerge which are not associated with the Boulware--Deser ghost and are intrinsic to the vielbein formulation. These new degrees of freedom are associated with the Lorentz transformations which encode the additional variables contained in the vielbein over the metric. Although they are not guaranteed to be ghostly, they are nevertheless infinitely strongly coupled on Minkowski spacetime and are not part of the spin-2 multiplet. Hence their existence implies the uniqueness of the Einstein--Hilbert term as the kinetic term for a massive graviton.}
\maketitle

\section{Overview}

The dynamics of a Lorentz Invariant massless spin 2 particle are remarkably constrained. Attempting to build a theory of an interacting massless spin 2 particle, one is led uniquely (up to second order in derivatives) to the Einstein--Hilbert action plus a cosmological constant \cite{Gupta:1954zz,Feynman:1996kb,Weinberg:1965rz,Deser:1969wk,Boulware:1974sr}. At higher order in derivatives, insisting that there are no new propagating degrees of freedom one is lead to the special Lovelock combinations \cite{Lovelock:1971yv,Zanelli:1999fs,Liu:2008zf}.
Remarkably some of these restrictions survive under a weaker set of assumptions. For example, without assuming Lorentz invariance (while maintaining space diffs), demanding ghost--freedom or more precisely only 2 propagating modes, one is led inevitably back to the Einstein--Hilbert action \cite{Khoury:2011ay,Khoury:2013oqa}.
\\

Recently there has been renewed interest in Lorentz--invariant massive gravity, which has recently found a ghost--free formulation \cite{deRham:2010ik,deRham:2010kj,Hassan:2011hr,Hassan:2011ea,Hassan:2011zd,deRham:2011rn,Hassan:2012qv}. For recent reviews see \cite{deRham:2014zqa,Hinterbichler:2011tt}.
The question arises, if by including a mass, we relax the assumption of a massless spin--2 particle, are we able to build new consistent derivative interactions? An affirmative answer to this question would have several important consequences. First, new ghost--free kinetic interactions, if they exist, should be included in the Lagrangian on general effective field theory grounds. Second, new kinetic terms could potentially allow for new cosmological solutions or different phenomenology.\\

The existence of potentially new derivative interactions \cite{Folkerts:2011ev,Hinterbichler:2013eza,Folkerts:2013mra} are also related to the existence of potentially new matter couplings studied recently \cite{deRham:2014fha,deRham:2014naa,Noller:2014sta,Hinterbichler:2015yaa,Matas}. Coupling matter to a composite vielbein or metric, is equivalent under a field redefinition to a minimal coupling of matter to the vielbein/metric with a non--standard kinetic term. Recently it was suggested that no new degrees of freedom arise nonlinearly if matter couples to the composite vielbein $e+ \alpha f$ \cite{Hinterbichler:2015yaa}. If correct, this would imply, by performing the field redefinition $e \rightarrow e- \alpha f$, that the following first order kinetic terms are ghost--free
\be
\int  \ep_{abcd}(e^a-\alpha f^a) \wedge (e^b-\alpha f^b) \wedge R^{cd}[\omega] \, ,
\ee
when matter is minimally coupled to gravity.
However in \cite{deRhamTolley} it is shown that this case is excluded due to the re--emergence of the BD ghost. \\

General Relativity (GR) and massive (multi--)gravity can be formulated in both the metric and vielbein language \cite{Nibbelink:2006sz,Chamseddine:2011mu,Hinterbichler:2012cn}.
In the vielbein the basic variables are $e^{\ a}_\mu$ and the spin connection $\om^{ab}_\mu$. There are two different formalisms: in second--order form, the spin connection is considered to be a known function of the vielbein, $\om=\om(e)$ (much like using the Christoffel connection in GR). In first order form, the spin connection is considered an independent variable with its own equation of motion (analogous to the Palatini formalism in GR). \\

For GR and for ghost--free massive gravity, these formalisms (metric and first-- and second-- order vielbein) are all equivalent.
However, in order to show the equivalence between the metric and vielbein formalisms, it is necessary to integrate out the additional variables encoded in the vielbein. In four dimensions, there are 6 additional variables which are encoded in the parameters of a Lorentz transformation. In standard massive gravity, these variables enter as non--dynamical auxiliary variables whose equations of motion
 impose the so--called symmetric vielbein (Deser--Van Nieuwenhuizen) condition \cite{Deser:1974cy,Hoek:1982za,Deffayet:2012zc}. This is most easily derived by introducing Lorentz \stu fields \cite{Ondo:2013wka} and deriving their equations of motion. The structure of these equations factors, as follows (see appendix~\ref{app:mGR} for more details)
 \ba
 \left(\text{Mass parameter dependent term} \right)  \times
 \left( \text{Symmetric vielbein condition}\right)=0 \, ,
 \ea
 where the symmetric vielbein condition is $\eta_{ab}\, e^a \wedge f^b =0$.
 If one modifies the kinetic structure then one modifies the equations for the Lorentz transformations (Lorentz \stu fields), and thus the symmetric vielbein conditon no longer holds. This alters the map between the vielbein and metric formalisms. This does not preclude the possibility of integrating out the additional variables to return to a metric formulation, but rather it implies is that it is no longer guaranteed that second order equations of motion for the vielbein imply second order equations for the metric and indeed this is one way to anticipate the problems we shall find. \\

In \cite{deRham:2014tga,Afshar:2014ffa}, it was shown in first order form that a set of related interactions in three dimensions contained a Boulware--Deser ghost. In what follows we shall extend the previous analyses by showing in four dimensions in first order form that the new kinetic terms inevitably lead to additional degrees of freedom. \\

In fact, in first order form a qualitatively new problem can arise. In massive gravity, the standard problem is the Boulware--Deser (BD) ghost \cite{Boulware:1973my}, see also \cite{Creminelli:2005qk,Deffayet:2005ys}. In the \stu language \cite{Siegel:1993sk,ArkaniHamed:2002sp}, the problem can be understood as follows. After restoring general covariance by introducing the four \stu fields $\phi^a$, one needs to see how many of the $\phi^a$ are dynamical fields. If all four  \stu fields become dynamical, the fourth will correspond to a BD mode. It will be infinitely strongly coupled around flat space (since if its kinetic term is present on flat space it is inevitably ghostly \cite{Fierz:1939ix}). \\

For new kinetic interactions in four dimensions expressed in the vielbein, new degrees of freedom beyond the BD ghost can emerge. In the vielbein language, we need to restore not just general covariance through the $\phi^a$ \stu fields, but also Local Lorentz invariance by introducing Lorentz \stu fields $\Lambda$. In vielbein form, the new danger that has no direct analogue in the metric formulation corresponds to giving kinetic terms to the Lorentz \stu fields, since $\Lambda$ is introduced with derivatives through the spin connection
\be
\om \rightarrow \Lambda^{-1} \om \Lambda + \Lambda^{-1} \d \Lambda \, .
\ee
This problem is compounded because there are a number of auxiliary fields in first order form. In four dimensions, there are 24 components of the spin connection $\om^{ab}_\mu$, but only 16 components of the vielbein $e^{\ a}_\mu$. This mismatch leads, in GR, to the need to integrate out several of the spin connection components to put the Hamiltonian in a canonical form. As we will see, GR is special in that it allows this procedure to be carried out without causing any of these auxiliary fields to become dynamical. \\

The obstructions to new kinetic terms in first order form go as follows:
\begin{itemize}
\item First, if there existed new healthy kinetic terms in first order form without any new degrees of freedom beyond that already present in massive gravity (or bi-- or multi--gravity), there should also be a healthy metric formulation since it should always be possible to integrate out or gauge fix all the additional degrees of freedom in the first order vielbein formalism. However in the metric language it has already been shown that no new kinetic terms could exist without introducing a ghost \cite{deRham:2013tfa}, (see also \cite{Matas} for further details). Because there is no metric formulation, there will always be an obstruction in converting the vielbein language to the metric language. As we will see below, this will take the form of new degrees of freedom.
\item  Even in the absence of coupling to matter, generically some number of the 6 Lorentz \stu degrees of freedom become dynamical. Hence this is a new problem, independent of the BD ghost. In unitary gauge the problem may be equivalently stated that the usual non--dynamical symmetric vierbein condition which removes these 6 degrees of freedom no longer holds, and gets replaced by a dynamical equation for some number of the 6 degrees of freedom. This problem is not visible in previous analyses. In the metric language in four dimensions considered in \cite{deRham:2013tfa}, the closest analogue is that the lapse and shift can appear with time derivatives. In the first order form in three dimensions considered in \cite{deRham:2014tga}, there are no auxiliary fields, and so there is only the standard BD ghost problem.
\end{itemize}

For the purposes of exposition we shall focus our consideration on the case of massive gravity with a Minkowski reference vielbein $f^a$. However all of our arguments will extend to the multi--gravity case, and since the massive gravity case is a clear decoupling limit of the multi--gravity then the absence of new kinetic terms in this limit is sufficient to rule out their existence in multi--gravity. Regardless, the problems we find with additional degrees of freedom can be translated almost verbatim to the general multi--gravity case. \\

We will focus on a particular well motivated kind of kinetic interaction
\be
\frac{\mpl^2}{4}\int \ep_{abcd} R[\om]^{ab} \wed \left( e^c \wed e^d + 2 \tau e^c \wed f^d + (\tau^2 + \kappa) f^c \wed f^d \right)\,,
\ee
where $R[\om]^{ab}$ is the curvature two--form. For $\kappa=0$, in the absence of coupling to matter, this is field redefinable to the Einstein--Hilbert term by $e \rightarrow e- \tau f$. Thus in the absence of matter, we expect any problem to be associated with $\kappa \ne 0$. This simple structure has many nice properties. They can be seen to arise from dimensional deconstruction \cite{deRham:2013awa} by choosing different procedures for discretizing the continuum action. The form structure guarantees the linearity of the Hamiltonian in the lapse and shift at least before integrating out any of the auxiliary fields, and this provides some hope for the absence of BD ghosts. We expect the problems we identify, however, to generically appear in modifications to the Einstein--Hilbert structure in first order form. Other kinds of interactions, without the wedge structure, will only have further opportunities for new degrees of freedom to arise and are thus not considered. \\

In what follows we shall present two different proofs of the existence of the ghost or more precisely additional degrees of freedom in these terms. Each presents a slightly different perspective on the problem.
\begin{enumerate}[label=\roman*.]
\item We shall first show that in the \stu language in the decoupling limit, new degrees of freedom can arise from the Lorentz \stu fields.
\item We then perform a full ADM analysis in the \stu language for perturbations around an arbitrary background, and show that the remaining theory has new degrees of freedom due simply to an enlarged phase space symplectic form. Since all constraints are first class there are none of the subtleties of the usual ghost--free proofs.
\end{enumerate}
Separately we sketch how the counting of the number of degrees of freedom works in unitary gauge and various other gauges.\\

The rest of the manuscript is organized as follows: In section~\ref{sec:New_Kinetic_Terms} we present the best candidates for new kinetic terms in first order form and sketch out the source of the problem. We then perform a decoupling limit analysis in section~\ref{sec:Decoupling_Limit} where the new degrees of freedom with their associated scales are emphasized. The counting of the degrees of freedom is also performed in other gauges in section~\ref{sec:Arbitrary_Background}. The full phase space counting is performed in section~\ref{sec:phaseSpace} where we show that the new kinetic terms leads to a least five additional configuration space degrees of freedom. We finally summarize our results in section~\ref{sec:outlook}.

To support our arguments we review how the symmetric vielbein condition arises in massive gravity in appendix~\ref{app:mGR} and we present the phase space counting in appendix~\ref{app:Counting}. Finally we present an explicit unitary gauge perturbative ADM analysis in appendix~\ref{sec:Perturbative_ADM}.

\section{New Kinetic Terms in first order form and origin of new degrees of freedom}
\label{sec:New_Kinetic_Terms}

In the absence of couplings to matter, any first order formulation which does not include new degrees of freedom must be equivalent to the metric formulation since it is always possible to integrate out or gauge fix the additional variables. For this reason the proof of absence of new kinetic terms in the metric formulation \cite{deRham:2013tfa} already rules out the absence of new kinetic terms (modulo coupling to matter) in the vielbein formalism.
Nevertheless, it is useful to perform an analysis directly in first order form to see where the problems arise. \\

Let us consider the following candidate new kinetic terms in four dimensions
\ba
\label{eq:NKT}
S = \frac{\mpl^2}{4}\int \ep_{abcd} R[\om]^{ab} \wed \left(e^a \wed e^b + 2 \tau e^a \wed f^b +
  \left(\tau^2 + \kappa\right)  f^a \wed f^b \right) \, ,
\ea
where $\tau$ and $\kappa$ are two dimensionless parameters. As already mentioned, any new kinetic term which does not preserve the form structure will break the linearity in the shift and the lapse already before integrating out the auxiliary variables and will only make the problems we face even worse. In this sense the candidate \eqref{eq:NKT} is the best hope for the new kinetic terms. Clearly if $\kappa=0$, this term is a linear field redefinition from the Einstein--Hilbert term and should be healthy at least in the absence of couplings to matter. Therefore any potential issue arising from \eqref{eq:NKT} should be solely carried by $\kappa$. This also means that there is no equivalent to the interaction carried by $\kappa$ in less than four dimensions.

\subsection{Equation of motion for $\om$}
First let us work in unitary gauge $f^a = 1^a$, (\ie  $f^{\ a}_{\mu}= \delta^{\ a}_{\mu}$). Then varying with respect to $\om$ yields the equation
\be
\label{eq:Eqspinconnection}
\left[\left( \d \hat{e}^a + \om^{as} \wed \hat{e}^s \right)\wed \hat {e}^b + \kappa \om^{as}\wed f^s \wed f^b\right]  - [a \leftrightarrow b]= 0 \, ,
\ee
where
\be
\hat{e}^a \equiv e^a + \tau f^a \, .
\ee
When $\kappa=0$, the theory is field redefinable to GR and we recover the normal torsion free condition. When $\kappa \ne 0$ we obtain a non--trivial modification to the torsion free condition leading to a quite different expression for $\om$. This difference will be the root of the problems. Solving for $\omega$ and substituting back into the action, we obtain a Lagrangian which contains terms quadratic in time derivatives of $e_i^{\ a}$ (with $i=1,\cdots,3$ and $a=0,\cdots,3$). \\

For $\kappa=0$, the kinetic term has a local Lorentz symmetry which implies that only 6 out of the 12 $e_i^{\ a}$ have an independent momentum conjugate. In the case of the modified kinetic term ($\kappa\ne0$), there is no obstruction to all 12 $e_i^{\ a}$ having a kinetic term leading to potentially 6 extra degrees of freedom and  we find that indeed all 12 $e_i^a$ have a kinetic term meaning that there is a significant enlargement of the phase space (see appendix~\ref{sec:Perturbative_ADM} for an explicit calculation). Note that this discussion is quite independent of the usual one of whether the lapse and shift continue to impose the constraint that removes the BD ghost.

\subsection{Re--emergence of ghosts: Sketch of the problem}

For massive gravity with a Minkowski reference metric, in the \stu formulation we can express the vielbein as
\be
f^{\ a}_{\mu} = \Lambda^a{}_b \partial_{\mu} \phi^b
\ee
or $f^a = \Lambda^a{}_b \d \phi^b$,  where $ \Lambda^a{}_b = (e^{\lambda})^a{}_b $ are the Lorentz \stu fields.
The problem with \eqref{eq:NKT} may now be seen in a number of different ways depending on the order of integrating out:
\begin{enumerate}
\item {\bf Integrating out $\omega^{ab}$ first:}\\
If we first chose to integrate out the spin connection $\omega^{ab}$, which enters quadratically and whose equation is algebraic, then it is necessary to first take the derivative off of $\omega^{ab} $ in $R[\om]^{ab} $ by integration by parts which will mean that both $e$ and $f $ pick up derivatives. The derivatives acting on $f$ give
\be
\d f^a = \d \Lambda^a{}_b \wedge \d \phi^b \, .
\ee
The resulting action will necessarily contain terms quadratic in $\d \Lambda^a_{\ b}$ which due to the form of the solution for $\om$ will not be simply wedge products. As such the resulting equation of motion for $\Lambda$ will contain derivatives. This implies that $\Lambda$ are no longer non--dynamical fields. Inevitably this will lead to additional degrees of freedom unless it can be shown that these higher derivative terms can be removed with a field redefinition, meaning that they are redundant operators. This is of course the special case $\kappa=0$.

\item {\bf Integrating out $\Lambda^{a}_{b}$ first:}\\
An analogous problem can be seen by performing this calculation in the opposite order. If we first derive the equation of motion for $\Lambda$ (which requires including the mass terms) then this gives the equation which usually determines the symmetric vielbein condition between the two vielbeins (see appendix \ref{app:mGR} for the standard analysis in massive gravity). However in the present case the solution for $\Lambda$ will depend on $R^{ab}(\om)$. As in the previous case this implies that the resulting action will now contain terms which are non--linear in derivatives of $\omega^{ab}$. Since we are already in first order form, even terms quadratic in time derivatives of $\omega^{ab}$ will imply the need to introduce an associated momentum conjugate and hence additional degrees of freedom. Once again the only loop hole in this argument is the possibility to do field redefinitions to remove these terms. As it turns out, this possibility only occurs when $\kappa=0$.

\item {\bf Imposing the Symmetric Vielbein condition from the outset:}\\
A different theory, but one that may also be considered is where we impose the symmetric vielbein condition from the outset as a constraint, namely
\be
\eta_{ab} e^a \wedge f^b =0 \, .
\ee
In four dimensions these are 6 equations which can be solved for the 6 Lorentz \stu fields which generates the square root structure. In fully \stu form $f^a$ will depend nontrivially on $ \d \phi^a$ through the square root. The problem, now is as before, integrating out the spin--connection will necessarily generate derivatives acting on $f^a$ and hence additional derivatives on the $\phi^a$ which are not of the form structure. Once again we would infer from these higher derivatives the existence of additional degrees of freedom, unless they are removable with a field redefinition. However in this case, we already fall under the umbrella of the proof given in \cite{deRham:2013tfa}.
\end{enumerate}

\subsection{Unitary gauge argument: Modification of symmetric vielbein condition}
If we work in unitary gauge, $f^a = 1^a$, then the problem of additional degrees of freedom can be understood as arising from the fact that the usual symmetric vielbein condition, which is a non--dynamical condition which fixed 6 of the $e_{\mu}^{\ a}$, becomes now a dynamical equation for at least some of those $6$. As we have already discussed, the symmetric vielbein condition is not a condition that should be imposed but is one that follows from the equations of motion. It is easiest to derive it by introducing the Lorentz \stu fields in $f^a$, and looking for their equation of motion. Expressed entirely in unitary gauge, the resulting generalization of the symmetric vielbein condition takes the form
\be
\eta_{ac} f_{\mu}^{\ c} \frac{\partial {\cal L}}{\partial f_{\mu} ^{\ b}} = \eta_{bc} f_{\mu}^{\ c} \frac{\partial {\cal L}}{\partial f_{\mu} ^{\ a}} \, .
\ee
Now in the absence of the new kinetic terms, only the mass term contributes to this equation and so we get a non--dynamical equation for 6 of the components of $e_{\mu}^{\ a}$.  In the presence of the new kinetic terms we see this equation includes terms linear in the curvature $R^{ab}(\omega)$. At the same time, the spin connection is no longer the usual GR one but is a modified function which will generically depend on time derivatives of all 12 $e_i^{\ a}$. That this is the case can be seen more easily in the \stu form where the solution for $\omega$ will explicitly depend on $\d \Lambda^a{}_b$. Regardless, this implies that generically $R(\omega)$ will contain double time derivatives of those 6 degrees of freedom (captured by  $\Lambda^a{}_b$) which were previously non dynamical. In other words the symmetric vielbein condition has now become a dynamical equation for some number of the 6 degrees of freedom it previously constrained.

\section{Decoupling limit of New Kinetic Term}
\label{sec:Decoupling_Limit}

\subsection{New contribution in the decoupling limit}

The problems stated  previously can already be seen in the decoupling limit of the above action. In the decoupling limit we choose to scale all the fields by their natural canonical normalization for fluctuations around Minkowski spacetime. The decoupling limit scaling is such that $R[\om] /m^2$ remains finite, \ie the Vainshtein radius remains finite. In four dimensions this corresponds to sending $m \rightarrow 0$, keeping $\Lambda^3  = m^2 \mpl$ fixed. One slight difference with the usual decoupling limit derivation is we must account for the fact that the form of the action remains the same under the replacement $e \rightarrow e+ c f$, with arbitrary parameter $c$. This leads to an apparent ambiguity in how we choose to perform the decoupling limit. This ambiguity is resolved by ensuring the recovery of the correct form of the action for fluctuations around Minkowski. \\

To anticipate this ambiguity lets us define $e = \hat e + c f$ where $c$ is at present arbitrary. In terms of $\hat e$ we choose to take the decoupling limit in terms of this vielbein for which the action is
\be
S = \frac{\mpl^2}{4}  \int \ep_{abcd} R[\om]^{ab} \wed \left(\hat e^c \wed \hat e^d + \alpha e^c \wed f^d+ \beta f^c \wed f^d \right)\,,
\ee
where $\alpha = 2(c+\tau)$, $\beta = (c+\tau)^2+\kappa$. We then proceed as usual by defining the canonically normalized fields \cite{Ondo:2013wka}
\ba
&& \omega^{ab} = \frac{1}{\mpl} \mu^{ab} \, , \\
&& \lambda_{ab} = \frac{1}{m \mpl} \hat \lambda_{ab}  \, , \\
&& \phi^a = x^a + \frac{1}{m \mpl} A^a + \frac{1}{\Lambda^3} \partial^a \pi  \, , \\
&& \hat e^a = 1^a + \frac{1}{\mpl}v^a \, .
\ea
In addition we should choose to scale the potentially new kinetic interaction parameters $\alpha $ and $\beta $ in such a way that the leading non--zero interaction arises from each term. The correct scaling that leads to a finite non--zero action from each term is
\ba
&& \alpha=m  \hat \alpha \, \\
&& \beta=m  \hat \beta \,.
\ea
The new terms that arise in the decoupling limit are then simply
\ba
S_{\rm new} &=&  \hat \alpha \int \ep_{abcd} \,  \d \mu^{ab} \wed 1^c \wed \left( \hat \lambda^{d}{}_e \(1^e+\frac{1}{\Lambda^3} \d \partial^e \pi\) + \d A^d \)  \\
&+& 2 \hat \beta \int \ep_{abcd} \,  \d \mu^{ab} \wed \(1^c+\frac{1}{\Lambda^3} \d \partial^c \pi\) \wed \left( \hat \lambda^{d}{}_e (1^e+\frac{1}{\Lambda^3} \d \partial^e \pi) + \d A^d \)\nn\,,
\ea
which on integration by parts gives
\ba
S_{\rm new} &=&   \hat \alpha \int \ep_{abcd} \,   \mu^{ab} \wed 1^c \wed \d \hat \lambda^{d}{}_e \wed \( 1^e+\frac{1}{\Lambda^3} \d \partial^e \pi \) \\
&+&2 \hat \beta \int \ep_{abcd} \,   \mu^{ab} \wed \(1^c+\frac{1}{\Lambda^3} \d \partial^c \pi\) \wed \d \hat \lambda^{d}{}_e \wed \( 1^e+\frac{1}{\Lambda^3} \d \partial^e \pi \) \, .\nn
\ea
As usual the decoupling limit can be chosen so that the fluctuations around Minkowski have the usual behavior. In this case we choose $\hat \alpha = - 2 \hat \beta$ in order to ensure no contribution from the new kinetic term at quadratic order. This fixes $c$ to be the solution of
\ba
2 (c+\tau) = - 2 (c +  \tau)^2-2 \kappa\,.
\ea
With this choice the new kinetic term is
\ba
S_{\rm new}  &=&-\hat \alpha\int \ep_{abcd} \,   \mu^{ab} \wed \(\frac{1}{\Lambda^3} \d \partial^c \pi\) \wed \d \hat \lambda^{d}{}_e \wed \( 1^e+\frac{1}{\Lambda^3} \d \partial^e \pi \) \, .
\ea
Taken together with the usual decoupling limit action, we may now integrate the spin--connection, which will clearly generate terms with time derivatives of $\hat\lambda^a_{\ b}$. However regardless of the value for $c$, as long as $\kappa \ne 0$, there is no choice for $c$ for which there is no contribution from these two terms in the decoupling limit.

\subsection{Phase space counting of decoupling limit theory}

In order to anticipate our future calculation, let us perform a phase space counting of the degrees of freedom already in the decoupling limit in this form. We refer to appendix~\ref{app:Counting} for the general phase space counting. \\

The decoupling limit symmetries are
\begin{itemize}
\item linear diffs {\bf (4)}
\be
v^a \rightarrow v^a + \d \xi^a\,.
\ee
\item linear local Lorentz transformations {\bf (6)}
\ba
&&\mu^{ab} \rightarrow \mu^{ab} + \d \chi^{ab} \, ,\\
&& v^a \rightarrow v^a + \chi^a{}_b 1^b \, ,
\ea
where $\chi^{ab}$ is antisymmetric.
\item linear $U(1)$ {\bf (1)}
\be
A \rightarrow A + \d \chi\,.
\ee
\end{itemize}

In the absence of the new kinetic terms the phase space form comes from the EH term expanded to second order
\be
S_{\rm EH} = \frac{\mpl^2}{4}\int 2 \epsilon_{abcd} \, 1^a \wed v^b \wed \d \mu^{cd} + \epsilon_{abcd} 1^a \wed 1^b \wed \mu^{ae} \wed \mu_{e}^{\ d} \, ,
\ee
as well as the contributions from the usual mass terms which provide the kinetic terms for $A^a $ and $\pi$ \cite{Ondo:2013wka}. Due to the manifest $U(1)$ symmetry, only 3 of the four $A^a$ are dynamical with 3 of the Lorentz \stu fields $\lambda^a_{\ b}$ acting as their conjugate momenta. $\pi$ on the other hand enters the decoupling limit action in this form with double time derivatives,
 and so we should at this stage introduce additional phase space variables of momenta $P_{\pi}$ conjugate to $\pi$ to reduce the action entirely to first order form. \\

The normal counting of phase space degrees of freedom proceeds as follows,
$A_0$, $\omega^{ab}_0$ and $v^a_0$ are the Lagrange multipliers which generate the above symmetries, the phase space is composed of 12 $v_i^a$ conjugate to 12 of the $\mu_i^{ab}$. 3 $A_i$ and their associated momentum conjugate which are 3 (the boosts) of the $\lambda^a_b$, 1 $\pi$ and its momentum conjugate $P_{\pi}$. 6 of the remaining $\mu_i^{ab}$ (those of the form $\mu_i^{jk}$ where $i$ and $k$ are distinct) and the remaining 3 Lorentz \stu fields (the rotations) $\lambda_{ab}$ are auxiliary fields which may be integrated out. Since all of the constraints are associated with symmetries, \ie are first class, {\it we are guaranteed that these auxiliary fields may be integrated out in a way which preserves the constraints}. This follows simply from the fact that we know the Lagrangian is invariant under the symmetry, and so their must be constraints that generate this symmetry which can be written entirely in terms of the phase space variables.
\\

As a result the final counting of phase space degrees of freedom is
\be
2 \times  (12 + 3 +1) - 2 \times ( 4+6+1) = 2 \times 5 \, .
\ee

Now with the addition of the new kinetic term, we have a contribution to the phase space measure that includes some components of $\lambda^a_b$ and the 6 $\mu_i^{jk}$ with $i \ne k$ in conjugate pairs. Since the rotation part of the Lorentz \stu fields and the  6 $\mu_i^{jk}$ were previous auxiliary, the phase space is now larger.  Specifically there appears to be at least $2 \times 3$ phase space degrees of freedom larger. However there are no additional symmetries in the system and so there appear to be 3 extra configuration degrees of freedom corresponding to the Lorentz rotations becoming dynamical:
\be
2 \times  (12 + 3 +1+3 + \dots) - 2 \times ( 4+6+1) = 2 \times (8 + \dots)\, .
\ee
Thus this is not a problem of the BD ghost (which has to do with the kinetic term for $\pi$), but rather distinct set of fields which are infinitely strong coupling around Minkowski spacetime. Indeed the BD ghost continues to be absent since this is associated with the $U(1)$ symmetry.

\section{Counting Degrees of Freedom in Different Gauges}
\label{sec:Arbitrary_Background}

The previous decoupling limit argument which indicated the existence of at least\footnote{We say `at least' since there may exist further degrees of freedom whose contribution vanishes in the decoupling limit.} 3 additional degrees of freedom coming from the Lorentz \stu fields can be extended to the full nonlinear theory. We shall see how this counting can arise from 4 points of view: 
\begin{enumerate}[label=(\alph*)]
\item a unitary gauge ADM analysis, 
\item a Lorentz \stu analysis, 
\item a diffeomorphism \stu analysis, 
\item and finally a full Lorentz plus diffeomorphism \stu analysis. 
\end{enumerate}
For reasons that will become clear the latter approach is most convenient, despite the large number of phase space variables,  in that no second class constraints arise.

\subsection{Unitary gauge counting of degrees of freedom}

If we work in unitary gauge, then the 16 components of the vierbein may be naturally split up into the four $e_{0}^{\ a}$, which encode the lapse and shift, and the 12 $e_i^{\ a}$. Starting from the general form of the new kinetic terms
\be
S = \frac{\mpl^2}{4}\int \ep_{abcd} R[\om]^{ab} \wed \left(e^c \wed e^d + 2 \tau e^c \wed f^d + \left(\tau^2 + \kappa\right)  f^c \wed f^d \right) \, ,
\ee
we may integrate out the spin--connection, whose equation of motion is algebraic, to reduce the kinetic term to something quadratic in derivatives of $e_i^{\ a}$. Schematically this will have the form
\be
S =  \frac{\mpl^2}{4}\int  \d^4 x  L^{ij}{}_{ab}\dot e_i^{\ a} \dot e_i^{\ b} + \dots
\ee
We show explicitly in Appendix~\ref{sec:Perturbative_ADM} that for $\kappa \ne 0$ the $12 \times 12$  matrix $ L^{ij}{}_{ab}$ has rank 12. This means that all 12 components of $e_i^{\ a}$ have a non--zero kinetic term. We may thus define a momentum conjugate $P^i_{\ a}$ to each of them and invert this relation to put the action in first order form
\be
S = \int \d^4x \, P^i_{\ a} \dot e_i^{\ a} - e_0^{\ a} H_a(e_i^{\ a},P^i_{\ a}) - W(e_i^{\ a},P^i_{\ a}) \, .
\ee
This could have been equivalently obtained by recognizing that $12$ of the components of the spin--connection were conjugate to 12 of the $e_i^{\ a}$ and the remaining 6 components can be integrated out as auxiliary fields. However crucially the form structure of the original theory ensures that the first order form is manifestly linear in $e_0^{\ a}$. Consequently we will have 4 second class primary constraints which naively reduce the phase space counting to
\be
2 \times (12) - 4  = 2  \times 10 \ ,
\ee
which are the usual 5 of the massive graviton and an additional $5$ out of the $6$ Lorentz boosts and rotations which encode the difference between the vielbein and metric. At this point is necessary to check whether the primary constraints lead to secondary constraints. If they do not then we have 5 additional degrees of freedom, rather than the 1 of the usual BD problem. If they do lead to secondary constraints then the maximum number of secondaries we can have is 4 leading to
\be
2 \times (12) - 2 \times 4  = 2  \times 8 \ .
\ee
This is the counting suggested by the decoupling limit. We may further worry about the potential existence of tertiary constraints, however we shall demonstrate below by different means that tertiary constraints never arise. Thus the minimum number of degrees of freedom is 8 which is 3 too many. This is consistent with the decoupling limit calculation.

\subsection{Lorentz \stu counting of degrees of freedom}

A slightly different counting can be achieved by introducing Lorentz \stu fields (but not diff \stu fields), by replacing $f^a$ with $\Lambda^a{}_b 1^b$. In this case it is easiest not to integrate out the the spin--connection but to keep it as part of the phase space.
The phase space symplectic form now depends on the 12 $e_i^{\ a}$, the 6 $\Lambda^a_{\ b}$ and the 18 $\omega_i^{ab}$. There thus appear to be $2 \times 18$ phase space degrees of freedom. In addition now we have 6 local Lorentz symmetries whose constraints are the coefficients of $\omega_0^{ab}$. Finally we still have $e_0^{\ a}$ generating 4 second class primary constraints. Thus the counting of degrees of freedom is
\be
2 \times 18 - 2 \times 6 - 4 = 2 \times 10
\ee
\ie $10=5+5$ degrees of freedom as before. Once again it is necessary to check whether the primary constraints lead to secondary ones. If all 4 do we would be left with 8 degrees of freedom.

\subsection{Diffeomorphism \stu counting of degrees of freedom}

The previously identified second class constraints can be made first class by introducing diffeomorphism \stu fields $\phi^a$. Let us now consider the form of the action in which $f^a = \d \phi^a$ so we have diffeomorphism invariance but not local Lorentz invariance. Working in first order form,  the total set of phase variables are the 12 $e_i^{\ a}$, the 4 $\phi^a $ and the 18 $\omega_i^{ab}$. By inspecting the symplectic form we see that 16 of the $\omega_i^{ab}$ will be momenta conjugate to the 16 $e_i^{\ a}$, $\phi^a $. This leaves 2 $\omega_i^{ab}$ non--dynamical. However these will be fixed by 2 of the second class constraints generated by $\omega_0^{ab}$ leaving behind $4$ second class constraints. The constraints generated by $e_0^{\ a}$ are now first class since they are just those associated with diffeomorphism invariance. The counting is then
\be
2 \times 12 + 2 \times 4 - 2 \times 4 - 4 = 2 \times 10 \, .
\ee
In this formulation it is the 4 remaining second class constraints associated with $\omega_0^{ab}$ that may lead to secondary constraints which once again reduce the system to at least 8 given the absence of tertiary constraints.

\subsection{Fully \stu formalism counting of degrees of freedom}

Finally let us see how this works in the fully `St\"uckelberg--ized' form for which $f^a = \Lambda^a_{\ b} \d \phi^b$ and we treat the spin--connection as an independent variable. The total  phase space symplectic form is build out of 18 $\omega_i^{ab}$, 12 $e_i^{\ a}$, 4 $\phi^a$ and 6 $\Lambda^a_{\ b}$. Assuming this symplectic form is invertible in this 40 dimensional phase space, and given that $e_0^{\ a}$ and $\omega_0^{ab}$ generate $4+6$ first class symmetries (diffs plus local Lorentz), so that there are no secondary constraints, then the counting is
\be
(18+12+4+6)-2 \times (4+6) = 2 \times 10 \, .
\ee
Thus the fully \stu version has the advantage over all the other approaches that there are no second class constraints to worry about. However we have assumed that the symplectic form in the 40 dimensional phase space is invertible. This is not necessarily the case and it could be that some number of the 40 variables do not enter independently in the symplectic form. At a minimum however we will see that all 18 $\omega_i^{ab}$ admit an independent momentum conjugate. This means that the phase space is at least 36 dimensional. Since all the constraints are first class we are again guaranteed that all auxiliary variables can be integrated out without spoiling the constraints. Thus the minimum number of degrees of freedom is
\be
2 \times 18 - 2 \times (4+6) = 2 \times 8 \, .
\ee
This confirms that there can be no tertiary constraints and is once again consistent with the result in the decoupling limit.

\section{Phase Space for New Kinetic Terms}
\label{sec:phaseSpace}

It is our claim that the new kinetic terms for a massive graviton lead to a minimum of 3 and generically 5 additional configuration space degrees of freedom, \ie a total of  $10=5+5$. We shall now confirm this as follows:
\begin{itemize}

\item We work in the fully \stu form in which $f^a = \Lambda^a_{\ b} \d \phi^b$.

\item We perturb around a specific background and confirm that there are at least 36 and generically 40 independent variables entering the phase space symplectic form.

\item Given the manifest local Lorentz and diff symmetries the total number of degrees of freedom around this background are at least $36-2 \times 10 = 2 \times 8 $ and generically $40-2 \times 10 = 2 \times 10 $, confirming the minimum 8 implied by the decoupling limit analysis but also showing that there are 2 additional degrees of freedom that do not arise in the decoupling limit.

\end{itemize}

To reiterate, we begin with the following fully `St\"uckelberg--ized' form of the action
\ba
S = \frac{\mpl^2}{4} \int \epsilon_{abcd}  \, R[\omega]^{ab} \wedge \Bigg[\left( e^c \wedge e^d + 2 \tau e^c \wedge \left( e^{\lambda} \right)^d_{\ f} \d \phi^f\right) \\
+ (\tau^2 +\kappa) \left( e^{\lambda} \right)^c_{\ f} \d \phi^f \wedge \left( e^{\lambda} \right)^d_{\ g} \d \phi^g\Bigg] \, .\nn
\ea
For simplicity we will switch off the mass terms since these can never cancel additional  degrees of freedom which may already arise from the kinetic term. \\

Without loss of generality, we choose to expand around a background which takes the following form in unitary gauge
\ba
&& e^a = \bar e^a + v^a \\
&& \omega^{ab} = \bar \omega^{ab} + \mu^{ab} \\
&& \phi^a = x^a + A^a \\
&& \lambda^a_{\ b} = 0 + \lambda^a_{\ b} \, .
\ea
Perturbing the action to quadratic order and focusing only on those terms which can enter in the phase space symplectic form, \ie those terms with at least one time derivative
\ba
S_{(2)} &=&  \frac{\mpl^2}{4} \int \epsilon_{abcd} \,  \d \mu^{ab} \wedge \Big[ 2 ( \bar e^c +\tau 1^c)  \wedge v^d  +2 \tau \bar e^c \wedge \lambda^d_{\ f} 1^f + 2 (\tau^2 +\kappa) 1^c \wedge \lambda^d_{\ f} 1^f    \nn \\
 &+& 2 \tau \bar e^c \wedge \d A^d + 2 (\tau^2 + \kappa) 1^c \wedge \d A^d \Big] \, \nn \\
&+&  \frac{\mpl^2}{4} \int \epsilon_{abcd} R[\bar \omega]^{ab} \wedge \Big[ 2 \tau \bar e^c \wedge  \lambda^d_{\ f} \d A^f + 2 ( \tau^2 +\kappa) 1^c \wedge \lambda^d_{\ f} \d A^f      \nn \\
 &+& 2 ( \tau^2 +\kappa) \lambda^c_{\ f} 1^f \wedge \d A^d     + (\tau^2 + \kappa) \d A^c \wedge \d A^d  \Big] + \text{non derivative terms} \, .
\ea
Note that the term $\d \mu^{ab} \wedge  2 (\tau^2 + \kappa) 1^c \wedge \d A^d $ is a total derivative and may be removed for the purposes of counting the number of degrees of freedom. \\

This full expression is still quite complicated, however since the counting of phase space degrees of freedom is local, it may be performed in the vicinity of any point $x$. In order to get a feel for the answer let us temporarily focus on a special case. We can always add a matter source\footnote{This could be achieved by looking at a background where $\bar{\omega}= \bar{\Lambda}^{-1} \d \bar \Lambda$ but here we are not even saying that the background has zero curvature since the condition \eqref{eq:R} is only required in the vicinity of a single point of spacetime.} so that in the vicinity of a single point $x$ the background satisfies
\be
\label{eq:R}
R[\bar \omega]^{ab}(x) =0 + {\cal O} (x^2) \, ,
\ee
then at a minimum the number of degrees of freedom will be determined by the first part
\ba
S'_{(2)} &=&  \frac{\mpl^2}{4} \int \epsilon_{abcd} \,  \d \mu^{ab} \wedge \Big[ 2 ( \bar e^c +\tau 1^c)  \wedge v^d  +2 \tau \bar e^c \wedge \lambda^d_{\ f} 1^f + 2 (\tau^2 +\kappa) 1^c \wedge \lambda^d_{\ f} 1^f     \nn \\
&+&   2 \tau \bar e^c \wedge \d A^d + 2 (\tau^2 + \kappa) 1^c \wedge \d A^d \Big] \,   .
\ea
Integrating by parts to remove the derivatives on $A^d$ and using the fact that $\d^2=0$
\ba
S'_{(2)} &=&  \frac{\mpl^2}{4} \int \epsilon_{abcd} \,  \d \mu^{ab} \wedge \Big[ 2 ( \bar e^c +\tau 1^c)  \wedge v^d  +2 \tau \bar e^c \wedge \lambda^d_{\ f} 1^f + 2 (\tau^2 +\kappa) 1^c \wedge \lambda^d_{\ f} 1^f     \nn \\
&+& 2 \tau  \d \bar e^c \wedge A^d  \Big] \,   .
\ea
Further keeping track of only those terms with time derivatives and integrating by parts to put all time derivatives on $\mu^{ab}$ this action reduces to
\be
S'_{(2)} = \int \d^4 x \, \partial_t  \mu_i^{ab} \,  P^i_{ab}(v_i^a, A^a, \lambda^a_{\ b}) + v_0^a C_a + A_0 C + \mu_0^{ab} C_{ab}\,,
\ee
where $P^i_{ab}(v_i^a, A^a, \lambda^a_{\ b}) $ are the 18 momenta conjugate to the 18 $\mu_i^{ab}$, and $C_a, C, C_{ab}$ the constraints. The `would be' $U(1)$ symmetry associated by with $C$ is broken by the non--derivative terms, in particular the mass terms, and so $C$ is not a true constraint. On the other hand $C_a$ and $C_{ab}$ are the first class constraints associated with linearized diffeomorphism and local Lorentz transformations.  \\

Now the crucial point is that the 18 $P^i_{ab}$ are a complicated background--dependent linear combination of the 22 variables $v_i^a, A^a, \lambda^a_{\ b}$. Furthermore these linear combinations depend on the background dependent variables $\bar e_i^{\ a}$ (12) and $(\d \bar e^{a})_{ij}$ (24).
It is thus straightforward to see that generically these 18 $P_i^{ab}$ are linearly independent. For instance, we already know from the results of Appendix \ref{sec:Perturbative_ADM} that all 12 $v_i^a$ are conjugate to 12 of the $\mu_i^{ab}$. For generic values of $(\d \bar e^a)_{ij}$ then the 4 $A^a$ may be made conjugate to 4 additional $\mu_i^{ab}$. We then require only that $\bar e^{\ a}_i$ may be chosen so that 2 of the $\lambda^a_{\ b}$ are conjugate to the remaining 2 linearly independent $\mu_i^{ab}$. Thus the phase space $(\mu_i^{ab}, P^i_{ab})$ has at least 36 dimensions. It is guaranteed that the remaining 4 of the $(v_i^a, A^a, \lambda^a_{\ b})$ can be integrated out without losing the constraints $C_a$ and $C_{ab}$ by virtue of the gauge invariance of the Lagrangian.

\subsection*{Special case $\kappa=0$}

Once again, the special case $\kappa=0$, $\tau \ne 0$ is exceptional, since then in the absence of matter couplings we know that the terms with $\lambda^a{}_b$ and $\d A^a $ can be removed with a field redefinition. It is easy to see that this is the only case for which this is true, and this is confirmed concretely in Appendix \ref{sec:Perturbative_ADM} where we find that the dimension of the phase space in unitary gauge changes as soon as $\kappa \ne 0$, see eq.~\eqref{det}, where no prior knowledge of the field redefinition is needed. \\

\subsection*{Special case $\tau=0$, $\kappa \ne 0$}

As an example of the more generic case we may consider the special choice $\tau=0$, $\kappa \neq 0$ for which
\ba
S'_{(2)} &=&  \frac{\mpl^2}{4} \int \epsilon_{abcd} \,  \d \mu^{ab} \wedge \( 2  \bar e^c   \wedge v^d  +  2 \kappa 1^c \wedge \lambda^d_{\ f} 1^f    \) \,  .
\ea
This is particular simple in that $A^a$ does not enter, only the 12+6 variables $e_i^{\ a}$ and $\lambda^a_{\ b}$. There is more than enough freedom in the 12 background functions $\bar e_i^{\ a}$ to ensure that all 12+6 variables enter independently and are conjugate to the 18 $\mu^{ab}_i$. Thus at least for perturbations around a point for which $R[\bar \omega]^{ab}(x) =0+ {\cal O}(x^2)$, it is the 4 $A^a$ which are auxiliary and may be integrated out.
\\

Let us now consider what happens when $ R[\bar \omega]^{ab} \ne 0 $ or when the contribution from the mass terms is taken into account.  Then it is easy to see that the full 40 dimensions of the phase space active. The salient point is the presence of the term $(\tau^2 + \kappa) \epsilon_{abcd} R[\bar \omega]^{ab} \wedge   \d A^c \wedge \d A^d$. This may be integrated by parts into $(\tau^2 + \kappa) \epsilon_{abcd} \d R[\bar \omega]^{ab} \wedge   \d A^c  A^d$ and for generic backgrounds for which $\d R[\bar \omega]^{ab} \ne 0$ implies that the previously auxiliary $A^a$ enter the phase space symplectic form, and from this term alone we see that effectively two of the $A^a$ act as momentum conjugate to the other 2. Hence the total phase space dimension is $40$ and so the final count is
\be
40 - 2 \times 10 = 2 \times 10 \, .
\ee
Although we just made this argument for $\tau=0$, it is clear that generically the same will be true for $\tau \ne 0$ since the number of degrees of freedom cannot decrease, except for special degenerate points at which the perturbations will be infinitely strong coupled. \\

To reiterate, from the point of view of the unitary gauge and other gauge calculations, what we have excluded is the possible existence of secondary and tertiary constraints. By working in full \stu form, we have promoted all constraints to be first class and hence avoided the need to perform a cumbersome Dirac constraint analysis.

\section{Outlook}
\label{sec:outlook}

At this point there are now a powerful set of results that indicate that the kinetic interactions of a massive spin--2 particle must take the form of the Einstein--Hilbert action (up to field redefinitions), even though the mass term explicitly breaks diffeomorphism invariance. In the metric language, a BD ghost was shown to be always present for kinetic terms other than the Einstein-Hilbert one in \cite{deRham:2013tfa}. \\

 In this paper we extend these considerations to the (unconstrained) vielbein formalism, and analyze the most promising candidate kinetic terms which are free of the BD ghost (in the absence of matter coupling).  For these candidates, we identified a qualitatively new source of unhealthy degrees of freedom separate from the standard BD ghost problem that have no metric counterpart. Specifically those degrees of freedom associated with the broken local Lorentz symmetry may become dynamical. For the choice of kinetic term we have considered generically 5 of the Lorentz boosts/rotations become dynamical. Although we have not exhausted the analysis, we fully anticipate that more general choices of kinetic terms will only be worse, potentially exciting all 6 Lorentz boosts/rotations and the BD ghost. Our results indicate that any consistent set of interactions for a massive graviton must have a representation in the metric language or fall afoul of the problems we have identified. \\
 
 Some number of these new degrees of freedom manifest themselves already in the $\Lambda$ decoupling limit, if the coefficients of the new kinetic terms are scaled appropriately (order unity times $m \mpl^2$). Then at best they can be considered within the context of an EFT expansion, where the cutoff is $\Lambda$, and as such they will spoil the standard realization of the Vainshtein mechanism. Alternatively, the coefficient of these operators may be chosen to be smaller, so that they do not arise in the decoupling limit, and hence the masses of the new degrees of freedom are above $\Lambda$, but then these new kinetic terms will be unimportant for any phenomenology.
 \\

These arguments extend to multi-gravity theories (and higher dimensions) where the problem multiples due to the increasing number of Lorentz \stu fields (and their number of components), \ie the increasing number of broken local Lorentz symmetries.

\section*{ Acknowledgments}
We would like to thank Kurt Hinterbichler, Nick Ondo and Rachel Rosen and for useful discussions. AJT is supported by Department of Energy Early Career Award DE-SC0010600. CdR is supported by a Department of Energy grant DE-SC0009946. AM is supported by an NSF-GRFP.

\appendix

\section{Review of symmetric vielbein condition in massive gravity}
\label{app:mGR}
In massive gravity with the Einstein--Hilbert kinetic term, the symmetric vielbein (Deser--Van Nieuwenhuizen) condition arises as a consequence of the equations of motion. This can be seen already in unitary gauge, and is due to the fact that the EH kinetic term is invariant under local Lorentz transformations, but the mass term is not. As a consequence varying the Lorentz part of the vielbein will give an equation which arises entirely from the mass term and hence is non--dynamical. For all choices of mass terms, this equation is solved by the symmetric--vielbein condition\footnote{There are also disconnected branches of solutions for which the symmetric--vielbein condition does not hold (see for example the discussion in \cite{Banados:2013fda}. However these branches generically have ghosts and are not continuously connected with the massive graviton perturbative vacuum, hence we disregard them from the outset.}. \\

To see this more clearly we follow the argument of \cite{Ondo:2013wka} which introduces explicit \stu fields for the local Lorentz symmetry.
We introduce the Lorentz \stu fields through the reference vielbein $f$ as
\be
f^a \rightarrow \Lambda^a{}_b f^b \, ,
\ee
then we can write the mass term in the generating function form
\be
\mathcal{L}_m = \sum_n c_n \left(\frac{1}{n!} \frac{\partial^n}{\partial \mu^n} \right) \det \left(e + \mu \Lambda f\right) \, .
\ee
The equation of motion for the Lorentz \stu field is then
\ba
\sum_n c_n \left( \frac{1}{n!} \frac{\partial^n} {\partial \mu^n} \right)  \mu \left[ {\rm det} (e + \mu \Lambda f) \right] {\rm Tr} \left[ \left( \delta \Lambda  f \right) \left( e - \mu \Lambda f \right)^{-1} \right] = 0 \, , \\
\sum_n c_n \left( \frac{1}{n!} \frac{\partial^n} {\partial \mu^n} \right) \mu \left[ {\rm det} (e + \mu \Lambda f) \right] {\rm Tr} \left[ \left( \delta \Lambda \Lambda^{-1} \eta \right) \eta (\Lambda f) \left( e - \mu \Lambda f \right)^{-1} \right] = 0 \, .
\ea
Because $\delta \Lambda  \Lambda^{-1} \eta$ is antisymmetric by virtue of the properties of Lorentz transformations, what it multiplies in the trace must be symmetric. We can show that this reduces to the $\mu$ independent equation
\be
(\Lambda f)^T \eta e = e^T \eta(\Lambda f) \, ,
\ee
which on returning to unitary gauge $\Lambda=1$ is just the symmetric vielbein condition
\be
\eta_{ab} e^a_\mu f^b_\nu =\eta_{ab}  f^a_\mu  e^b_\nu \, .
\ee
Note that the crucial property that the variation $\delta \Lambda$ factors from the rest of the mass term, leading to a $\mu$ independent condition (which is therefore independent of the parameters in the potential).

\section{Phase space counting in first order form}
\label{app:Counting}

In this appendix we review the standard phase space counting for (massive) gravity in first order form with the usual EH kinetic term. The action for standard massive gravity in $D=4$ dimensions is
\be
S = \frac{\mpl^2}{4} \int \ep_{abcd}\left( R[\om]^{ab} \wed e^c \wed e^d + m^2 (c_1 e^a \wed e^b \wed e^c \wed f^d + \cdots) \right) \, ,
\ee
where $f$ is a fixed reference vielbein, and where
\be
R[\om]^{ab} = \d \om^{ab} + \om^{ac} \wed \om^{cb}\,.
\ee
In first order form we treat $\om^{ab}$ and $e^a$ as independent fields.
It is important to keep track of how many of each field is present: in 4 dimensions there are 16 $e^{\ a}_\mu$ fields and 24 $\om^{ab}_\mu$ fields.

\subsection{Phase space measure for perturbations around a background in locally inertial frame}

For the purpose of understanding the kinetic structure it will be useful to consider the Einstein--Hilbert kinetic structure perturbed to quadratic order around an arbitrary background.
More precisely, we perturb
\ba
&& e^a_\mu = \bar{e}^a_\mu + v^a_\mu \nn \\
&& \om^{ab}_\mu = \bar{\om}^{ab}_\mu + \mu^{ab}_\mu\,.
\ea
We work in a locally inertial frame (LIF), where in the vicinity of a point $x$ we may use local diffs and Lorentz to impose $\bar{e}^{\ a}_\mu=\delta^a_\mu$. Note that even though the vielbein is diagonal, we should allow the derivatives of the vielbein, and the background spin connection, to be arbitrary.
Then the Einstein Hilbert term gives rise to the following symplectic structure
\ba
\ep_{abcd} \bar{D} \mu^{ab} \wed v^c \wed \bar{e}^d &\rightarrow & \ep_{abcd} \ep^{ijk} \dot{\mu}^{ab} v^c_j \delta^d_k \nn \\
&=& \dot{\mu}^{IJ}_i \left( \ep^{ijk} \ep_{IJk} v^0_j \right) + \dot{\mu}^{I0}_i \left( \ep^{ijk} \ep_{IJk} v^J_j \right) \nn \\
&=& \dot{\mu}^{i j}_i v_j^0 + \left( \dot{\mu}^{k0}_k \delta^j_i - \dot{\mu}^{j0}_i \right)v^i_j\,,
\ea
where $D$ is the covariant derivative with respect to the background.
There are 18 $\mu^{ab}_k$ fields. Of them, 12 appear in the phase space measure: all 9 fields of the form $\mu^{i0}_j$, as well as 3 `trace fields' $\mu^{ij'}_i$. The remaining 6 auxiliary fields are of the form $\mu^{ij}_k$, where $i$ and $k$ are distinct. Thus the trick of going to a LIF allows us to cleanly identify who are the non--dynamical parts of the spin--connection.

\subsection{Counting for GR}
As a warm up, let us first consider the counting when we set $m=0$.
The symplectic form is determined by
\be
\Omega = \ep_{abcd} \d \omega^{ab} \wed e^c \wed e^d\,.
\ee
Since there are only 12 $e_i^{\ a}$'s, then only 12 of the $\omega_i^{ab}$ independently contribute to the symplectic form, the remaining 6 are non--dynamical, \ie auxiliary fields which should be integrated out.
Then there are
\ba
(e^{\ a}_i, \om^{ab}_i) \rightarrow 2\times 12\ \text{naive phase space dofs}\,.
\ea
In addition, there are some first class constraints associated with diffs and LLTs
\ba
&& v^a_0 \ \text{generate 4 first class constraints} \nn \\
&& \om^{ab}_0\ \text{generate 6 first class constraints}
\ea
The form structure guarantees that it is the 0 components that enter without time derivatives and act as lagrange multipliers.
Finally, there are the 6 fields of the form $\om^{ij}_k$ that are auxiliary and as mentioned can be integrated out.  As a result, the final counting is then
\be
24 - 2\times(4+6)=2 \times 2\ \text{ phase space dofs}
\ee
which is correct for GR.

\subsection{Integrating out auxiliary spin--connection and Lorentz constraint}

In the above counting we used the fact that 6 of the components of the spin--connection $\omega^{ij}_k$, are non--dynamical and can be integrated out without spoiling the existence of the 6 constraints associated with the Lagrange multipliers $\omega_0^{ab}$. Putting this another way, the constraints associated with $\omega_0^{ab}$ truly constrain the components of the spin--connection which are in the phase space measure. For the Einstein--Hilbert term this may be easily verified by going again in LIF. In order to determine wihich components of the spin--connection are fixed by the 6 $\omega_0^{ab}$  it is sufficient to look at the quadratic term
\be
\epsilon_{abcd} \, e^a \wedge e^b \wedge \omega^{cf} \wedge \omega_f^{\ d} \, .
\ee
Concentrating on the part linear in $\omega_0^{ab}$ we have
\be
\epsilon_{abc} \epsilon^{ijk} e_i^a e_j^b \omega_k^{cl} \omega_0^{l0} - \epsilon_{abc}\epsilon^{ijk} e_i^a e_j^b \omega_k^{0l} \omega_0^{lc} + \dots
\ee
Then in the locally inertial frame $e_i^{\ a} = \delta_i^a$ we find
\be
2 \omega_k^{k l} \omega_0^{l0} - 2 \omega_k^{0l} \omega_0^{lk } + \dots\,.
\ee
Thus varying with respect to $\omega_0^{l0}$ imposes a condition on $ \omega_k^{k l}$ and varying with respect to $\omega_0^{lk } $ imposes a condition on $\epsilon_{ilk} \omega_k^{0l} $. Since both of these sets of fields is already in the phase space measure, then as promised the constraints truly fix 6 of the phase space variables and the additional components of $\omega_i^{jk}$ can be safely integrated out.

\subsection{Counting for massive gravity in the \stu language}

We can now generalize this to standard massive gravity. We will introduce diff $\phi^a$ and LLT $\Lambda^{ab}$ \stu fields in the reference vielbein $f$, \ie $f^a \rightarrow \Lambda^a{}_b d \phi^b$. The phase space is now enlarged since the mass term gives a contribution to the symplectic term that comes from derivatives acting on $\phi^a$ and will include dependence on $e^{\ a}_i$ and $\Lambda^a{}_b$. For the ghost--free mass terms, some combination of $\phi^a$ will not be dynamical. \\

We thus have a {\bf naive} phase space
\ba
(e^{\ a}_i, \om^{ab}_i) &\rightarrow&  2 \times 12 \nn \\
(\phi^a, \Lambda^{ab}) &\rightarrow& 2 \times 4\,.
\ea
There are 6 $\om^{ab}_i$ as before and 2 $\lambda^{ab}$ that do not appear in the phase space and can be eliminated using their non--dynamical equation of motion. Crucially, this can be done without introducing new degrees of freedom.

Then there are first class constraints
\ba
v^a_0 &\rightarrow& 4\ {\rm constraints} \nn \\
\mu^{ab}_0 &\rightarrow& 6\ {\rm constraints}\,.
\ea
At this point the counting is
\be
2\times(12+4) - 2\times(4+6) = 2\times (5+1)\,,
\ee
corresponding to massive gravity and a BD ghost. \\

However, in the case of the ghost--free mass terms, in fact only 3 of the 4 momenta conjugate to $\phi^a$ are independent.  Consequently there are an additional pair of second class constraints which remove the BD ghost. In practice this means that only 3 of the 6 $\Lambda^a_b$, specifically the boosts, act as momenta conjugate to the 3 dynamical $\phi^a$. The remaining 3 rotations are auxiliary variables which may be integrated out.

\section{Unitary Gauge Perturbative ADM analysis}
\label{sec:Perturbative_ADM}

\subsection{Auxiliary variables}
In this appendix we show explicitly that as soon as  $\kappa\ne 0$ the size of the unconstrained phase space in unitary gauge is increased from 12 to 24. This is the root of the problem of additional degrees of freedom in unitary gauge. We will utilize a perturbative approach which is sufficient to prove the point.
When working in unitary gauge $f^{\, a}_\mu=\delta^{\, a}_\mu$, it is useful  to start with the same formalism as that developed by Hinterbichler and Rosen in \cite{Hinterbichler:2012cn} and express the vielbein as a Lorentz boost on an upper triangular matrix,
\ba
e^{\ a}_\mu=\(
\begin{array}{cc}
N \gamma +N^i e_i^{\ a} v_a & N v^a+N^i e_i^{\ b}\(\delta_b^a+\frac{1}{\gamma+1}v_b v^a\) \\
e_i^{\ a}v_a & e_i^{\ b}\(\delta_b^a+\frac{1}{\gamma+1}v_b v^a\)
\end{array}\)\,,
\ea
where $\gamma$ is the Lorentz factor $\gamma=\sqrt{1+v^2}$.
The 16 components of the vielbein are split into the lapse $N$, the three shift $N^i$, the three boost vectors $v^i$ and the nine spatial components of the vielbein $e_i^{\, j}$.
The spatial vielbein can also further be split into six symmetric vielbein and three rotations -- or three antisymmetric vielbein. \\

In this formalism we keep the spin--connection as independent and its equation of motion is given in \eqref{eq:Eqspinconnection}. The strategy is the following: 1. Integrate out the spin--connection. We can then check explicitly that the resulting Lagrangian for the kinetic term involve no time derivatives, neither on the lapse nor on the shift. Moreover the kinetic term is linear in the  shift and has the correct dependence on the lapse which ensures the primary constraint that removes the BD ghost. However this is not sufficient, we then proceed  by 2. Determining how many of the vielbein are dynamical. For that we can compute the rank of the following twelve--dimensional matrix
\ba
\mathbb{L}_{IJ}=\frac{\p^2\L}{\p \dot e_{I}\p \dot e_{J}}\,,
\ea
where we use the notation $\{e_{I}\}_{I=1,\cdots,12} \equiv \{e^{\ a}_i\}_{i=1,\cdots,3 \ a=0,\cdots,3}$. For GR and standard massive gravity, the rank of this matrix is six, which means that six of the vielbein are auxiliary variables (these are precisely the boosts and rotations). For the candidate new kinetic term introduced we will see that the rank of this matrix is larger than six as soon as $\kappa\ne 0$, which implies that some of the Lorentz \stu are dynamical and the phase space is increased. Since these new additional degrees of freedom are absent at leading order about Minkowski they correspond to infinitely strongly coupled degrees of freedom. \\

This approach is in principle straight--forward but in what follows we solve for the spin--connection perturbatively and define the perturbed vielbein as follows
\ba
N=1+\e\, \delta N,\qquad N^i=0+ \e\, \delta N^i, \qquad v^i=0+\e\, \delta v^i, \qquad e^{\ i}_j=\delta^{\ i}_j+\e\, \delta e^{\ i}_j\,,
\ea
where we introduced the parameter $\e$ as a way to keep track of the order in perturbations.
We will see that at leading and cubic order both the boosts and the rotations (which perturbatively are captured by the antisymmetric part of the spatial vielbein) remain auxiliary variables.

\subsection{Kinetic Matrix}

We proceed by integrating out the spin--connection and deriving the kinetic matrix $\mathbb{L}_{IJ}$. To simplify the procedure, we diagonalize the matrix at leading order and denote by $U$ the diagonalization matrix. Then, up to quadratic order in $\e$, the kinetic matrix takes the form
\ba
\mathbb{D}_{IJ} =  U^{-1}\, \mathbb{L}_{IJ}\, U =
\(\begin{array}{cc}
\bar A &  0 \\
0 & 0
\end{array}\)
+\e
\(\begin{array}{cc}
\delta_1 A & \delta_1 B\\
\delta_1 C & 0
\end{array}\) +\e^2
\(\begin{array}{cc}
\delta_2 A & \delta_2 B\\
\delta_2 C & \delta_2 D
\end{array}\) +\mathcal{O}(\e^3)\,,
\ea
where all the blocks are $6 \times 6$ matrices and where $\bar A$ is diagonal
\ba
\label{eq:eigenvaluesA}
\bar A= \frac{4(1+\tau )^2}{\kappa + (1+\tau )^2}\, {\rm diag}\(-2,1,1,1,1,1\)\,.
\ea
The negative eigenvalue is associated with the conformal mode which would be a ghost if it was not projected out by the Hamiltonian constraint.

To leading order (\ie zeroth order in $\e$ in the kinetic matrix), $\mathbb{L}_{IJ}$ is clearly of rank six and the directions associated with its vanishing eigenvalues represent the auxiliary variables. They are nothing other than the three boosts $v^i$ and the rotations, which are captured here by the antisymmetric contributions to the spatial vielbein, $a_{ij}=\delta e^{\ i}_j- \delta e^{\ j}_i$.
Not only are the $v^i$ and $a_{ij}$ not dynamical but they actually fully disappear from the whole kinetic structure (at least at that order) and the only place they enter is in the mass term from where they impose the symmetric vielbein condition. Therefore at zeroth order in $\e$ the vielbein and metric formulation of this theory are the same. Similarly at first order in $\e$  the kinetic matrix has still six vanishing eigenvalues. \\

The situation is however dramatically different at higher order in perturbations. At quartic order in $\e$ in the action, or second order in $\e$ in the kinetic matrix, we may  compute its determinant, using the fact that
\ba
\det  \mathbb{L}_{IJ}= \e^{12}\(\det\bar A\) \det\(\delta_2 D-\delta_1 C\ \bar A^{-1}\ \delta_1 B \)+\mathcal{O}(\e^{13})\,.
\ea
An explicit calculation shows that
\ba
\label{det}
\det\(\delta_2 D-\delta_1 C\ \bar A^{-1}\ \delta_1 B \)=-32 \(\frac{\kappa^{2}}{(\kappa+(1+\tau)^2)^3}\)^6\times \\
\times \left[\mathcal{E}_{ijk}\delta e^\ell_{\ 0}\delta e^k_{\ 0}\delta e^m_{\ 0}\(\delta e^i_{\ n}\delta e^j_{\ \ell}\delta e_m^{\ \, n}
+\delta e_n^{\ \, i}\delta e_\ell^{\ j}\delta e^n_{\ m}\)\right]^2\ne 0\,,\nn
\ea
where in the last expression all the indices are spacelike and are raised and lowered with respect to $\delta_{ij}$.
The first thing to notice is that if $\kappa=0$ then the result vanishes no matter what $\tau$ is. This is simply a reflection of the fact that no matter what $\tau$ is (so long as $\kappa=0$), the kinetic term is just a field redefinition away from the Einstein--Hilbert one and the formalism developed here is fully able to adapt for this without needing to set it by hand. However as soon as $\kappa=0$, $\det  (\mathbb{L}_{IJ})\ne 0$  and all the twelve vielbein $\delta e_i^{\ a}$ turn alive already at quartic order in perturbations.

\bibliographystyle{JHEPmodplain}
\bibliography{refs}

\end{document}